\documentclass{jpsj2}

\title{%
Asymptotic Behavior of the Conductance in Disordered Wires
with Perfectly Conducting Channels
}

\author{%
Yositake {\sc Takane}
}

\inst{%
Department of Quantum Matter, Graduate School of Advanced Sciences of Matter,
Hiroshima University, Higashi-Hiroshima 739-8530, Japan
}

\recdate{ \hspace{50mm} }

\abst{%
We study the conductance of disordered wires with unitary symmetry
focusing on the case in which $m$ perfectly conducting channels are present
due to the channel-number imbalance between two-propagating directions.
Using the exact solution of the Dorokhov-Mello-Pereyra-Kumar (DMPK) equation
for transmission eigenvalues, we obtain
the average and second moment of the conductance in the long-wire regime.
For comparison, we employ the three-edge Chalker-Coddington model
as the simplest example of channel-number-imbalanced systems with $m = 1$,
and obtain the average and second moment of the conductance
by using a supersymmetry approach.
We show that the result for the Chalker-Coddington model
is identical to that obtained from the DMPK equation.
}

\kword{%
unitary class, DMPK equation, Chalker-Coddington model, supersymmetry
}

\begin{document}
\sloppy
\maketitle

\section{Introduction}

The statistical property of electron transport in a disordered
quantum wire is independent of microscopic details of the system, and is
mainly determined by the symmetries the system possesses.~\cite{beenakker1}
In ordinary disordered systems, only time-reversal and spin-rotation
symmetries play a relevant role.
According to the presence or absence of these two symmetries,
disordered quantum wires are classified into either of the three standard
universality classes (i.e., orthogonal, unitary and symplectic).
The orthogonal class consists of systems having both time-reversal and
spin-rotation symmetries, while the unitary class is characterized
by the absence of time-reversal symmetry.
The systems having time reversal-symmetry
without spin-rotation symmetry belong to the symplectic class.
It has been believed that
although details of transport properties differ from class to class,
Anderson localization inevitably arises in all the three standard classes
(i.e., the conductance decays exponentially with increasing sample length $L$
and eventually vanishes in the limit of $L \to \infty$).
However, this widely accepted understanding dose not always hold true.
Recent studies on the symplectic universality class show that if the number of
conducting channels is odd, one channel becomes perfectly conducting without
backward scattering.~\cite{ando,takane1,takane2,takane3,sakai1,sakai2,sakai3}
Due to the presence of this perfectly conducting channel,
the dimensionless conductance behaves as $g \to 1$ with increasing $L$,
and thereby Anderson localization disappears.
In contrast to this, such an anomalous behavior does not arise
in the ordinary case with an even number of conducting channels.
Thus, we must separate the symplectic class into two subclasses of
symplectic-even and symplectic-odd according to whether the number of
conducting channels is even or odd.~\cite{takane1,sakai2}

The realization of the symplectic-odd class revives attention to
disordered wire systems with the channel-number imbalance
between two-propagating directions.
More than a decade ago, Barnes, Johnson and Kirczenow~\cite{barnes} pointed out
that if the number of conducting channels in one propagating direction
is by $m$ greater than that in the opposite direction,
the dimensionless conductances $g$ for the majority direction and
$g'$ for the minority direction satisfy $g = g' + m$.
Performing a numerical simulation, they obtained evidence that
$g \to m$ and $g' \to 0$ in the long-$L$ limit.
This indicates that $m$ perfectly conducting channels are present
only in the majority direction.
Note that the channel-number imbalance leads to
the breaking of time-reversal symmetry,
so that their argument applies only to the unitary class.
Recently, Hirose, Ohtuski and Slevin~\cite{hirose} have proposed
the Chalker-Coddington model~\cite{chalker} with an odd-number of edge channels
and studied its electron transport properties.
The Chalker-Coddington model can be viewed as a stacking of
alternating left-moving and right-moving chiral edge channels with random
tunneling between adjacent ones.~\cite{lee}
It should be noted that in the odd-edge case, the channel number in one
propagating direction is by one greater than that in the opposite direction.
This situation is equivalent to that in the case of $m = 1$
considered in ref.~\citen{barnes}.
Performing a numerical simulation, they also confirmed that
$g \to 1$ and $g' \to 0$ in the long-$L$ limit.

One may think that the channel-number imbalance is
rather unrealistic in actual systems.
However, Wakabayashi {\it et al.}~\cite{wakabayashi} have recently shown that
such a system can be realized in zigzag nanographene ribbons.~\cite{fujita}
Inspired by this observation, the present author and Wakabayashi~\cite{takane4}
have formulated a random-matrix theory for the unitary universality class
with the channel-number imbalance.
Let us focus on a disordered wire system of length $L$ having
$N + m$ left-moving channels and $N$ right-moving channels.
In this case, $m$ left-moving channels become perfectly conducting
and the dimensionless conductances $g$ and $g'$ for the left-moving
and right-moving channels, respectively, satisfy $g = g' + m$.
They have derived the Dorokhov-Mello-Pereyra-Kumar (DMPK) equation, which
describes the evolution of the probability distribution for transmission
eigenvalues with increasing $L$,~\cite{dorokhov,mello1}
and analyzed the asymptotic behavior of $g' = g - m$ in the long-$L$ regime
by using an approximate method proposed by Pichard.~\cite{pichard}
They have shown that the localization length $\xi$,
which characterizes the exponential decay of $\exp [\langle \ln g' \rangle]$,
depends on $m$ as $\xi = 2Nl/(m+1)$,
where $l$ is the mean free path for the left-moving channels.
That is, $\xi$ decreases with increasing the number $m$ of
perfectly conducting channels.
This means that the presence of perfectly conducting channels suppresses $g'$.
Furthermore, they have also shown that the average and second moment of $g'$
behave as $\langle g' \rangle \sim {\rm e}^{-\frac{s}{4N}}$
for $m = 0$, $\langle g' \rangle \sim {\rm e}^{-\frac{ms}{N}}$ for $m \ge 1$,
$\langle {g'}^{2} \rangle \sim {\rm e}^{-\frac{(m+1)^{2}s}{4N}}$
for $0 \le m \le 3$, and
$\langle {g'}^{2} \rangle \sim {\rm e}^{-\frac{2(m-1)s}{N}}$ for $m \ge 3$,
where $s \equiv L/l$.
To examine these results in terms of an independent approach,
the present author and Wakabayashi~\cite{takane5} have studied the averaged
conductance $\langle g' \rangle$ in the three-edge Chalker-Coddington model,
i.e., the simplest nontrivial example of the channel-number-imbalanced
unitary class with $m = 1$.
They have found that $\langle g' \rangle \sim {\rm e}^{-\frac{s}{N}}$,
which is consistent with the DMPK result.
Although this supports the validity of the DMPK approach,
a more detailed comparison is highly desirable to deeply understand
the nature of this peculiar universality class.

In this paper, we study the average and second moment of the dimensionless
conductance $g'$ for the channel-number-imbalanced unitary class.
First, we obtain the asymptotic forms of $\langle g' \rangle$ and
$\langle {g'}^{2} \rangle$ in the long-$L$ regime for $0 \le m \le 4$
by using the existing exact solution~\cite{akuzawa} of the DMPK equation.
The exact solution, which is available only
in the unitary class,~\cite{beenakker2} enables us to obtain
the full asymptotic forms including a pre-exponential factor.~\cite{frahm1}
Second, we employ the $M$-edge Chalker-Coddington model for the cases of
$M = 2$ and $M = 3$,
and obtain $\langle g' \rangle$ and $\langle {g'}^{2} \rangle$ in the long-$L$
regime by using a supersymmetry approach.~\cite{mathur}
The $M = 2$ case is the simplest example of the ordinary unitary class,
while the $M = 3$ case is that of the channel-number-imbalanced unitary class.
We obtain the full asymptotic forms of $\langle g' \rangle$
and $\langle {g'}^{2} \rangle$ including a pre-exponential factor.
We show that the resulting asymptotic forms are
identical to those obtained from the DMPK equation.
This strongly supports the validity of the DMPK approach.

In the next section, we introduce the DMPK equation for
the channel-number-imbalanced unitary class and introduce its exact solution.
We obtain the average $\langle g' \rangle$ and second moment
$\langle {g'}^{2} \rangle$ of the dimensionless conductance
$g'$ in the long-$L$ regime using the exact solution.
In \S 3, we introduce the $M$-edge Chalker-Coddington model
and obtain $\langle g' \rangle$ and $\langle {g'}^{2} \rangle$
for the cases of $M = 2$ and $M = 3$.
Section 4 is devoted to summary.

\section{DMPK Approach}
We summarize the random-matrix theory for the channel-number-imbalanced
unitary class.~\cite{takane4}
We consider the case in which the number of left-moving channels is $N + m$,
while that of right-moving channels is $N$.
In this case, we can show that $m$ transmission eigenvalues in the left-moving
channels become unity.~\cite{barnes,hirose,takane4}
This indicates the presence of perfectly conducting channels.
If the set of the transmission eigenvalues for the right-moving channels is
$\{T_{1}, T_{2}, \dots, T_{N}\}$, that for the left-moving channels is
expressed as $\{T_{1}, T_{2}, \dots, T_{N}, 1, \dots, 1 \}$,
where we have identified the $N+1$ to $N+m$th channels as
the perfectly conducting ones.
The dimensionless conductance $g$ for the left-moving channels is given by
$g = \sum_{a=1}^{N+m}T_{a} = m + \sum_{a=1}^{N}T_{a}$,
while that for the right-moving channels is $g' = \sum_{a=1}^{N}T_{a}$.
It is easy to observe that $g = g' + m$.
We consider the behavior of $g' = g - m$ as a function of
the normalized system length $s \equiv L/l$,
where $l$ is the mean free path for the left-moving channels.
It should be noted that the mean free path $l'$
for the right-moving channels is not equal to $l$ due to the presence
of perfectly conducting channels only in the left-moving channels.
Indeed, we find that $l' = (N/(N+m))l$.
We define $\lambda_{a} \equiv (1-T_{a})/T_{a}$
and introduce the probability distribution $P(\{\lambda_{a}\};s)$
for the transmission eigenvalues.
The Fokker-Planck equation for $P(\{\lambda_{a}\};s)$,
which is usually called the DMPK equation, is expressed as~\cite{takane4}
\begin{equation}
      \label{eq:dmpk-lambda}
 \frac{\partial P(\{\lambda_{a}\};s)}{\partial s} =
   \frac{1}{N}\sum_{a=1}^{N} \frac{\partial}{\partial \lambda_{a}}
   \left( \lambda_{a}(1+\lambda_{a}) J
          \frac{\partial}{\partial \lambda_{a}}
            \left( \frac{P(\{\lambda_{a}\};s)}{J} \right)
   \right) 
\end{equation}
with
\begin{equation}
      \label{eq:jacob}
  J = \prod_{c=1}^{N} \lambda_{c}^{m}
      \times
      \prod_{b=1}^{N-1}\prod_{a=b+1}^{N}|\lambda_{a}-\lambda_{b}|^{2} .
\end{equation}
As stressed in ref.~\citen{takane4}, the factor
$\prod_{c=1}^{N} \lambda_{c}^{m}$ in $J$ represents the repulsion
arising from the $m$-fold degenerate perfectly conducting eigenvalue.
This reduces the non-perfectly conducting eigenvalues
$\{T_{1}, T_{2}, \dots, T_{N}\}$.
It should be mentioned that the equivalent DMPK equation was
proposed by Akuzawa and Wadati~\cite{akuzawa} in a rather formal context,
but they did not study electron transport properties.

The DMPK equation has been solved exactly for the ordinary case of
$m = 0$.~\cite{beenakker2,frahm1}
The exact solution for an arbitrary $m$ has been obtained
in ref.~\citen{akuzawa}.
In our notation, the probability distribution is given by
\begin{align}
   P(\{\lambda_{a}\};s)
   = \frac{1}{N!}{\rm Det}\{ K(\lambda_{a},\lambda_{b};s)\}_{a,b = 1,2,\dots,N}
\end{align}
with
\begin{align}
   K(\lambda,\lambda';s)
   = \sum_{j=1}^{N} Q_{j}(\lambda,s)h_{j}(\lambda',s) .
\end{align}
Here,
\begin{align}
      \label{eq:def_Q}
   Q_{j}(\lambda,s)
  & = \lambda^{m}G_{j-1}(m+1,m+1;-\lambda)
      {\rm e}^{-\frac{(2j+m-1)^{2}}{4N}s} ,
      \\
      \label{eq:def_h}
   h_{j}(\lambda,s)
  & = \int_{0}^{\infty} {\rm d}k L_{j}(k)c_{m}^{2}(k)
      {\rm e}^{-\frac{k^{2}}{4N}s}
      F\left(\frac{m+1-{\rm i}k}{2}, \frac{m+1+{\rm i}k}{2}, m+1; -\lambda
       \right) ,
\end{align}
where $G_{j-1}(m+1,m+1;-\lambda) \equiv
F(-j+1,m+j,m+1;-\lambda)$ is the Jacobi polynomial and
\begin{align}
   L_{j}(k)
   & = \prod_{l=1(l \neq j)}^{N}
       \frac{k^{2} + (2l+m-1)^{2}}{-(2j+m-1)^{2} + (2l+m-1)^{2}} ,
            \\
   c_{m}(k)
   & = \frac{1}{\sqrt{4\pi}}
       \frac{\Gamma\left(\frac{m+1+{\rm i}k}{2}\right)
             \Gamma\left(\frac{m+1-{\rm i}k}{2}\right)}
            {\Gamma(m+1)|\Gamma({\rm i}k)|} .
\end{align}
Note that $K(\lambda,\lambda';s)$ satisfies
\begin{align}
   & \int_{0}^{\infty} {\rm d}\lambda K(\lambda,\lambda;s) = N ,
             \\
   & \int_{0}^{\infty} {\rm d}\lambda'' K(\lambda,\lambda'';s)
                 K(\lambda'',\lambda';s) = K(\lambda,\lambda';s) .
\end{align}
Using these equations, we obtain the eigenvalue density
\begin{align}
   R_{1}(\lambda_{1};s)
 & = N \int_{0}^{\infty} {\rm d}\lambda_{2}\cdots{\rm d}\lambda_{N}
                       P(\{\lambda_{a}\};s)
        \nonumber \\
 & = K(\lambda_{1},\lambda_{1};s) ,
\end{align}
and the two-point correlation function
\begin{align}
   R_{2}(\lambda_{1},\lambda_{2};s)
 & = N(N-1) \int_{0}^{\infty} {\rm d}\lambda_{3}\cdots{\rm d}\lambda_{N}
                       P(\{\lambda_{a}\};s)
        \nonumber \\
 & = K(\lambda_{1},\lambda_{1};s)K(\lambda_{2},\lambda_{2};s)
     - K(\lambda_{1},\lambda_{2};s)K(\lambda_{2},\lambda_{1};s) .
\end{align}
We focus on the dimensionless conductance $g'$ for
the right-moving channels, in terms of which the dimensionless conductance $g$
for the left-moving channels is given by $g = m + g'$.
Using the eigenvalue density, we express the averaged dimensionless
conductance as
\begin{align}
        \label{eq:average}
   \langle g' \rangle
    = \int_{0}^{\infty} {\rm d}\lambda 
          \frac{1}{1+\lambda}R_{1}(\lambda;s) .
\end{align}
Using the eigenvalue density and the two-point correlation function,
we express the second moment as
\begin{align}
        \label{eq:2nd_moment}
   \langle {g'}^{2} \rangle
   =    \int_{0}^{\infty} {\rm d}\lambda
        \frac{1}{(1+\lambda)^{2}}R_{1}(\lambda;s)
      + \int_{0}^{\infty} {\rm d}\lambda \int_{0}^{\infty} {\rm d}\lambda'
        \frac{1}{(1+\lambda)(1+\lambda')}R_{2}(\lambda,\lambda';s) .
\end{align}
Alternatively, once $\langle g' \rangle$ is given, $\langle {g'}^{2} \rangle$
is obtained by using the scaling relation
\begin{align}
        \label{eq:scaling}
   N \frac{\partial \langle g' \rangle}{\partial s}
 = - \langle {g'}^{2} \rangle - m \langle g' \rangle ,
\end{align}
which can be derived from eq.~(\ref{eq:dmpk-lambda}) (see Appendix A).
Using these equations, we obtain $\langle g' \rangle$ and
$\langle {g'}^{2} \rangle$ in the asymptotic regime of $s \equiv L/l \gg 4N$.
It should be mentioned that the ordinary case of $m = 0$ has been
analyzed in ref.~\citen{frahm1}.

We obtain the asymptotic form of $\langle g' \rangle$.
We rewrite eq.~(\ref{eq:average}) as
\begin{align}
        \label{eq:average_rev}
   \langle g' \rangle
   = \int_{0}^{\infty} {\rm d}\lambda 
          \frac{1}{1+\lambda} \sum_{j=1}^{N}Q_{j}(\lambda,s)h_{j}(\lambda,s) .
\end{align}
From eq.~(\ref{eq:def_Q}), we observe that the term with $j = 1$ dominates
the others in the asymptotic regime of $s \gg 4N$,
so that we can neglect the terms with $j \ge 2$.
Equation~(\ref{eq:average_rev}) is then reduced to
\begin{align}
     \label{eq:basic-equation}
   \langle {g'}_{m} \rangle
  & = \int_{0}^{\infty} {\rm d}\lambda \frac{\lambda^{m}}{1+\lambda}
      \int_{0}^{\infty} {\rm d}k L_{1}(k)c_{m}^{2}(k)
      {\rm e}^{-\frac{k^{2}+(m+1)^{2}}{4N}s}
            \nonumber \\
  & \hspace{30mm}\times
      F\left(\frac{m+1-{\rm i}k}{2}, \frac{m+1+{\rm i}k}{2}, m+1; -\lambda
       \right) ,
\end{align}
where $G_{0}(m+1,m+1;-\lambda) = 1$ has been used.
Here and hereafter, we explicitly show the number of
perfectly conducting channels as a subscript.
Before describing the evaluation of eq.~(\ref{eq:basic-equation}),
we present the final results,
\begin{align}
     \label{eq:g_m=0}
   \langle {g'}_{0} \rangle
  & = \frac{\pi^{\frac{3}{2}}}{4}a_{N}
            \frac{1}{\sqrt{\frac{s}{4N}}^{3}} {\rm e}^{-\frac{s}{4N}} ,
          \\
     \label{eq:g_m=1}
   \langle {g'}_{1} \rangle
  & = \frac{2}{\pi^{\frac{3}{2}}}a_{N}
            \frac{1}{\sqrt{\frac{s}{4N}}} {\rm e}^{-\frac{s}{N}} ,
          \\
     \label{eq:g_m=2}
   \langle {g'}_{2} \rangle
  & = \frac{1}{8}a_{N} {\rm e}^{-\frac{2s}{N}} ,
          \\
     \label{eq:g_m=3}
   \langle {g'}_{3} \rangle
  & = \frac{1}{6\pi}a_{N} {\rm e}^{-\frac{3s}{N}} ,
          \\
     \label{eq:g_m=4}
   \langle {g'}_{4} \rangle
  & = \frac{9}{512}a_{N} {\rm e}^{-\frac{4s}{N}} ,
\end{align}
where
\begin{align}
   a_{N}
  = \frac{\pi}{4}L_{1}(0)
  = \frac{\Gamma\left(N + \frac{1}{2}\right)^{2}}{N!(N-1)!} .
\end{align}
Note that $a_{1} = \pi/4$ and $\lim_{N \to \infty}a_{N} = 1$.
We observe that the exponential decay of $\langle {g'}_{m} \rangle$
becomes faster with increasing $m$.
This should be attributed to the eigenvalue repulsion
arising from the $m$-fold degenerate perfectly conducting eigenvalue.

We briefly describe the derivation of eqs.~(\ref{eq:g_m=0})-(\ref{eq:g_m=4}).
For the cases of $m = 0$ and $m = 1$, we can exchange the order of
the integrations over $\lambda$ and $k$ in eq.~(\ref{eq:basic-equation})
and find
\begin{align}
        \label{eq:average'}
   \langle {g'}_{m} \rangle
    = \int_{0}^{\infty} {\rm d}k L_{1}(k)c_{m}^{2}(k)
      {\rm e}^{-\frac{k^{2}+(m+1)^{2}}{4N}s} I_{m}(k)
\end{align}
with
\begin{align}
   I_{m}(k)
    = \int_{0}^{\infty} {\rm d}\lambda \frac{\lambda^{m}}{1+\lambda}
      F\left(\frac{m+1-{\rm i}k}{2}, \frac{m+1+{\rm i}k}{2}, m+1; -\lambda
       \right) .
\end{align}
Adapting the method presented in ref.~\citen{frahm2},
we analytically obtain $I_{m}(k)$ as
\begin{align}
        \label{eq:I_0}
   I_{0}(k) & = \frac{\pi}{\cosh \left(\frac{\pi k}{2}\right)} ,
                  \\
        \label{eq:I_1}
   I_{1}(k) & = \frac{2\pi}{k \sinh \left(\frac{\pi k}{2}\right)} .
\end{align}
We next carry out the integration over $k$ in eq.~(\ref{eq:average'}).
The main contribution comes from the small-$k$ region of
$k \lesssim \sqrt{4N/s}$, and therefore we approximate as
$L_{1}(k) = L_{1}(0)$, $c_{0}^{2}(k) = (\pi/4)k^{2}$,
$c_{1}^{2}(k) = (4\pi)^{-1}k^{2}$, $I_{0}(k) = \pi$ and $I_{1}(k) = 4/k^{2}$.
After the $k$-integration,
we finally obtain eqs.~(\ref{eq:g_m=0}) and (\ref{eq:g_m=1}).
For $m \ge 2$, we employ a different approach,
which is applicable to the cases of $m \ge 1$.
We consider eq.~(\ref{eq:basic-equation}) in the large-$s$ limit.
Expecting that the main contribution to the $\lambda$-integration
comes from the region of $\lambda \gg 1$, we replace the hypergeometric
function in eq.~(\ref{eq:basic-equation}) by its asymptotic form
\begin{align}
     \label{eq: asymp_hgf}
   F\left(\frac{m+1-{\rm i}k}{2}, \frac{m+1+{\rm i}k}{2}, m+1; -\lambda
    \right)
        = \frac{\Gamma(m+1)\Gamma({\rm i}k)}
               {\Gamma\left(\frac{m+1+{\rm i}k}{2}\right)^{2}}
          \lambda^{- \frac{m+1-{\rm i}k}{2}} + {\rm c.c.}
\end{align}
in the large-$\lambda$ limit.
We treat the case of $m = 1$ as an example.
Note that the main contribution to the $k$-integration comes from the region of
$k \lesssim \sqrt{4N/s}$, in which we can approximate as
$L_{1}(k) = L_{1}(0)$ and $c_{1}^{2}(k) = (4\pi)^{-1}k^{2}$,
and eq.~(\ref{eq: asymp_hgf}) with $m = 1$ is reduced to
\begin{align}
   F\left(\frac{2-{\rm i}k}{2}, \frac{2+{\rm i}k}{2}, 2; -\lambda
    \right)
   = 2 \lambda^{-1} \frac{\sin\left(k\frac{\ln\lambda}{2} \right)}{k} .
\end{align}
After the $k$-integration, we obtain
\begin{align}
      \label{eq:g_m=1_expression}
   \langle {g'}_{1} \rangle
    = \frac{L_{1}(0)}{8\sqrt{\pi}}\frac{1}{\sqrt{\frac{s}{4N}}^{3}}
      {\rm e}^{-\frac{s}{N}}
      \int_{0}^{\infty} {\rm d}\lambda \frac{1}{1+\lambda}
      \frac{\ln\lambda}{2}
      \exp \left( -\frac{N}{s} \left(\frac{\ln\lambda}{2}\right)^{2} \right) .
\end{align}
Changing the variable from $\lambda$ to $x$ defined by
$\lambda = (\cosh(2x)-1)/2$,
we obtain
\begin{align}
      \int_{0}^{\infty} {\rm d}\lambda \frac{1}{1+\lambda}
      \frac{\ln\lambda}{2}
      \exp \left( -\frac{N}{s} \left(\frac{\ln\lambda}{2}\right)^{2} \right)
  \approx
        2 \int_{0}^{\infty} {\rm d}x \, x {\rm e}^{-\frac{N}{s}x^{2}}
  = \frac{s}{N} .
\end{align}
Substituting this into eq.~(\ref{eq:g_m=1_expression}) and using
$L_{1}(0) = (4/\pi)a_{N}$, we again arrive at eq.~(\ref{eq:g_m=1}).
Adapting the above method to the cases of $m \ge 2$,
we obtain eqs.~(\ref{eq:g_m=2})-(\ref{eq:g_m=4}).

We turn to the evaluation of the second moment in the asymptotic regime.
The results are summarized as follows:
\begin{align}
        \label{eq:g^2_m=0}
   \langle {g'}_{0}^{2} \rangle
  & = \frac{\pi^{\frac{3}{2}}}{16}a_{N}
            \frac{1}{\sqrt{\frac{s}{4N}}^{3}} {\rm e}^{-\frac{s}{4N}} ,
          \\
        \label{eq:g^2_m=1}
   \langle {g'}_{1}^{2} \rangle
  & = \frac{1}{4\pi^{\frac{3}{2}}}a_{N}
            \frac{1}{\sqrt{\frac{s}{4N}}^{3}} {\rm e}^{-\frac{s}{N}} ,
          \\
        \label{eq:g^2_m=2}
    \langle {g'}_{2}^{2} \rangle
  & = \frac{\pi^{\frac{3}{2}}}{128}a_{N}
            \frac{1}{\sqrt{\frac{s}{4N}}^{3}} {\rm e}^{-\frac{9s}{4N}} ,
         \\
        \label{eq:g^2_m=3}
    \langle {g'}_{3}^{2} \rangle
  & = \frac{1}{3\pi^{\frac{3}{2}}}a_{N}
            \frac{1}{\sqrt{\frac{s}{4N}}} {\rm e}^{-\frac{4s}{N}} ,
         \\
        \label{eq:g^2_m=4}
    \langle {g'}_{4}^{2} \rangle
  & = \frac{3}{128}a_{N} {\rm e}^{-\frac{6s}{N}} .
\end{align}
Again, we observe that the decay of $\langle {g'}_{m}^{2} \rangle$ becomes
faster with increasing $m$ due to the eigenvalue repulsion from
the $m$-fold degenerate perfectly conducting eigenvalue.

We briefly present the derivation of
eqs.~(\ref{eq:g^2_m=0})-(\ref{eq:g^2_m=4}).
From eq.~(\ref{eq:scaling}), the second moment is expressed in terms of
the averaged conductance as
\begin{align}
        \label{eq:scaling_rev}
   \langle {g'}_{m}^{2} \rangle
 = - N \frac{\partial \langle {g'}_{m} \rangle}{\partial s}
   - m \langle {g'}_{m} \rangle .
\end{align}
Substituting eqs.~(\ref{eq:g_m=0}) and (\ref{eq:g_m=1}) into this,
we straightforwardly obtain eqs.~(\ref{eq:g^2_m=0}) and (\ref{eq:g^2_m=1}),
respectively.
However, the right-hand side of eq.~(\ref{eq:scaling_rev}) vanishes if
we substitute eqs.~(\ref{eq:g_m=2})-(\ref{eq:g_m=4}).
This means that the second moment for $m \ge 2$ is related to
the next leading order correction to $\langle {g'}_{m} \rangle$.
Instead of evaluating such a correction, we directly obtain
$\langle {g'}_{m}^{2} \rangle$ using eq.~(\ref{eq:2nd_moment}).
In the asymptotic regime, eq.~(\ref{eq:2nd_moment}) is approximately reduced to
\begin{align}
        \label{eq:basic_2nd_moment}
   \langle {g'}_{m}^{2} \rangle
  & =   \int_{0}^{\infty} {\rm d}\lambda
        \frac{\lambda^{m}}{(1+\lambda)^{2}}
        \int_{0}^{\infty} {\rm d}k L_{1}(k)c_{m}^{2}(k)
        {\rm e}^{-\frac{k^{2}+(m+1)^{2}}{4N}s}
            \nonumber \\
  & \hspace{30mm}\times
        F\left(\frac{m+1-{\rm i}k}{2}, \frac{m+1+{\rm i}k}{2}, m+1; -\lambda
         \right) .
\end{align}
When $m \le 3$, we can exchange the order of
the $\lambda$- and $k$-integrations and find
\begin{align}
   \langle {g'}^{2} \rangle
    = \int_{0}^{\infty} {\rm d}k L_{1}(k)c_{m}^{2}(k)
      {\rm e}^{-\frac{k^{2}+(m+1)^{2}}{4N}s} \tilde{I}_{m}(k)
\end{align}
with
\begin{align}
   \tilde{I}_{m}(k)
    = \int_{0}^{\infty} {\rm d}\lambda \frac{\lambda^{m}}{(1+\lambda)^{2}}
      F\left(\frac{m+1-{\rm i}k}{2}, \frac{m+1+{\rm i}k}{2}, m+1; -\lambda
       \right) .
\end{align}
Adapting the method presented in ref.~\citen{frahm2},
we can analytically obtain $\tilde{I}_{m}(k)$ as
\begin{align}
   \tilde{I}_{2}(k) & = \frac{2\pi}{\cosh \left(\frac{\pi k}{2}\right)} ,
       \\
   \tilde{I}_{3}(k) & = \frac{12\pi}{k \sinh \left(\frac{\pi k}{2}\right)} .
\end{align}
After carrying out the integration over $k$,
we obtain eqs.~(\ref{eq:g^2_m=2}) and (\ref{eq:g^2_m=3}).
This approach cannot be applied when $m \ge 4$,
so we adapt the method used to derive eqs.~(\ref{eq:g_m=2})-(\ref{eq:g_m=4}).
It is applicable to the cases of $m \ge 3$.
We replace the hypergeometric function in eq.~(\ref{eq:basic_2nd_moment})
by its asymptotic form and integrate over $k$.
For $m = 4$, we obtain
\begin{align}
   \langle {g'}_{4}^{2} \rangle
    = \frac{3\sqrt{\pi}L_{1}(0)}{4^{5}}\frac{1}{\sqrt{\frac{s}{4N}}^{3}}
      {\rm e}^{-\frac{25s}{4N}}
      \int_{0}^{\infty} {\rm d}\lambda
      \frac{\lambda^{\frac{3}{2}}}{(1+\lambda)^{2}}
      \frac{\ln\lambda}{2}
      \exp \left( -\frac{N}{s} \left(\frac{\ln\lambda}{2}\right)^{2} \right) .
\end{align}
After the $\lambda$-integration,
we finally arrive at eq.~(\ref{eq:g^2_m=4}).

\section{Chalker-Coddington Model}

We consider an electron system consisting of $M$ chiral edge channels
as shown in Fig.~1.
\begin{figure}[btp]
\begin{center}
\includegraphics[height=6cm]{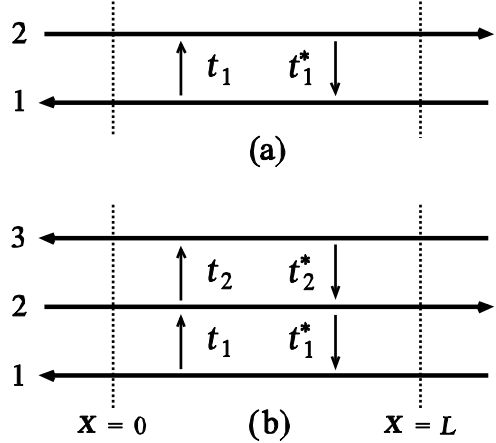}
\end{center}
\caption{Schematic sketch of the Chalker-Coddington model
for (a) the two-edge case of $M = 2$ and (b) the three-edge case of $M = 3$.
}
\end{figure}
The propagating direction of the $j$th edge channel is left (right)
if $j$ is odd (even), and each edge is coupled with adjacent
counter-propagating edge(s) by random tunneling.
Note that this system is essentially equivalent to
the Chalker-Coddington model.
We assume that the electron wavefunction $\phi_{j}(x)$
for the $j$th edge channel with energy $\varepsilon$ obeys~\cite{lee,mathur}
\begin{align}
 (-1)^{j}(-{\rm i})\partial_{x} \phi_{j}(x)
      + t_{j}(x)\phi_{j+1}(x) + t_{j-1}^{*}(x)\phi_{j-1}(x)
      = \varepsilon \phi_{j}(x) ,
\end{align}
where $\phi_{0}(x) = \phi_{M+1}(x) \equiv 0$ has been assumed.
Here, $t_{j}(x)$ and $t_{j}^{*}(x)$ represent the tunneling amplitude from
the $j$th edge to $j+1$th edge and that for the reverse process, respectively.
We assume that $\{t_{j}(x)\}$ are random variables
in the disordered region of $0 < x < L$, and vanish outside this region.
We regard $L$ as the length of our system, and the clean regions of $x \le 0$
and $L \le x$ play a role of the left and right electrodes, respectively.
The random amplitudes in the disordered region are assumed to be Gaussian
distributed with zero means.
That is, $\langle t_{j}(x) \rangle = 0$,
$\langle t_{j}(x) t_{j'}(x') \rangle = 0$ and
\begin{align}
      \label{eq:tunnel-el}
 \langle t_{j}(x) t_{j'}^{*}(x') \rangle = D\delta_{j,j'}\delta(x-x') ,
\end{align}
where $\langle \cdots \rangle$ represents the ensemble average.
We hereafter set $\varepsilon = 0$ without loss of generality.

The dimensionless conductance $g$ for the left-moving channels is
expressed as $g = \sum_{i,j= {\rm odd}}\mathcal{T}_{ij}$,
where $\mathcal{T}_{ij}$ is the transmission probability for an electron
incoming from the $j$th channel in the right electrode
and outgoing to the $i$th channel in the left electrode.
In a manner similar to this, $g'$ for the right-moving channels is expressed as
$g' = \sum_{i,j= {\rm even}} \mathcal{T}_{ij}$.
The dimensionless conductances satisfy $g = g'$
in the ordinary case of an even $M$.
In the odd-$M$ case, which is first studied by Hirose, Ohtuski and
Slevin,~\cite{hirose} the number of the left-moving channels is by one
greater than that of the right-moving channels,
so that one left-moving channel becomes perfectly conducting.
In this case, we observe that $g = 1 + g'$.
We consider the two-edge case of $M = 2$ and the three-edge case of $M = 3$
in the following.
In the notation used in the previous section, the former and latter correspond
to the cases of $N = 1$ with $m = 0$ and $N = 1$ with $m = 1$, respectively.
The three-edge case of $M = 3$ is the simplest nontrivial example
of the disordered wires with the channel-number imbalance.
Hereafter, we again explicitly show the number $m$ of
perfectly conducting channels as a subscript of $g'$.
We focus on the average and second moment of ${g'}_{m}$
($m = 0, 1$) as a function of $L$.
Note that ${g'}_{m} = \mathcal{T}_{22}$ in the two- and three-edge cases,
in which only one right-moving channel exists.

Mathur has shown for the $M = 2$ case that~\cite{mathur,note}
\begin{align}
  \langle {g'}_{0} \rangle
     = \frac{\pi^{\frac{5}{2}}}{16}\frac{1}{\sqrt{\frac{DL}{4}}^{3}}
       {\rm e}^{-\frac{DL}{4}} 
\end{align}
in the asymptotic regime of $DL \gg 1$.
This result is identical to eq.~(\ref{eq:g_m=0}) in the case of
$N = 1$ if we replace $DL$ by $s$.
This indicates that $D^{-1}$ should be identified
with the mean free path $\tilde{l}$ for the $M = 2$ case.
In the $M = 3$ case, each left-moving edge channel is directly coupled
to one right-moving edge channel as in the case of $M = 2$,
so the mean free path for the left direction is equal to $\tilde{l}$
under the condition of eq.~(\ref{eq:tunnel-el}).
However, because the right-moving channel interacts with
the two left-moving channels,
the corresponding mean free path becomes half of $\tilde{l}$.
Thus, for comparison with the argument in \S 2, in which the system length is
normalized by the mean free path for the left-moving channels,
we hereafter adopt $D^{-1}$ as the unit length scale common to
both the cases of $M = 2$ and $3$.
This is simply achieved by the replacement $DL \to s$.

We obtain the asymptotic forms of $\langle {g'}_{0}^{2} \rangle$,
$\langle {g'}_{1} \rangle$ and $\langle {g'}_{1}^{2} \rangle$
by adapting the supersymmetry approach presented by Mathur.~\cite{mathur}
The supersymmetry approach enables us to express the average and second moment
of the transmission probability in the from of a correlation function
for a non-random interacting fermion-boson system on a one-dimensional chain.
For the two-edge (three-edge) case,
we need to consider the two-site (three-site) chain which we describe below.
Let $c_{i \sigma}$ ($c_{i \sigma}^{\dagger}$) be the fermion annihilation
(creation) operator, and $b_{i \sigma}$ ($b_{i \sigma}^{\dagger}$)
be the boson annihilation (creation) operator, where $i = 1, 2, 3$ represents
the site number and $\sigma = \uparrow, \downarrow$.
The sites are one-to-one correspondence with
the edge channels in the original model.
The Hamiltonian $H$ for the two-edge case is identical to $H_{12}$
which represents the interaction between the 1st and 2nd sites.
The Hamiltonian for the three-edge case is given by $H = H_{12} + H_{23}$,
where $H_{23}$ represents the interaction between the 2nd and 3rd sites.
The explicit form of $H_{12}$ is given by
$H_{12} = H_{12}^{\rm F} + H_{12}^{\rm B} + H_{12}^{\rm FB}$ with
\begin{align}
  H_{12}^{\rm F}
         = & - \sum_{\sigma = \uparrow, \downarrow} 
               \left( c_{1 \sigma}^{\dagger}c_{1 \sigma} - 1/2
               \right)
               \left( c_{2 \sigma}^{\dagger}c_{2 \sigma} - 1/2
               \right)
             + c_{1 \uparrow}^{\dagger}c_{1 \downarrow}^{\dagger}
               c_{2 \uparrow}^{\dagger}c_{2 \downarrow}^{\dagger}
             + c_{1 \downarrow}c_{1 \uparrow}c_{2 \downarrow}c_{2 \uparrow} ,
   \\
  H_{12}^{\rm B}
         = &   \sum_{\sigma = \uparrow, \downarrow } 
               \left( b_{1 \sigma}^{\dagger}b_{1 \sigma} + 1/2
               \right)
               \left( b_{2 \sigma}^{\dagger}b_{2 \sigma} + 1/2
               \right)
             - b_{1 \uparrow}^{\dagger}b_{1 \downarrow}^{\dagger}
               b_{2 \uparrow}^{\dagger}b_{2 \downarrow}^{\dagger}
             - b_{1 \downarrow}b_{1 \uparrow}b_{2 \downarrow}b_{2 \uparrow} ,
   \\
      \label{eq:H_{FB}}
  H_{12}^{\rm FB}
         = & \sum_{\sigma = \uparrow, \downarrow}
               \left(
                 - c_{1 \sigma}^{\dagger}b_{1 \sigma}
                   c_{2 \sigma}^{\dagger}b_{2 \sigma}
                 - c_{1 \sigma}b_{1 \sigma}^{\dagger}
                   c_{2 \sigma}b_{2 \sigma}^{\dagger}
                 + c_{1 \sigma}^{\dagger}b_{1 \bar{\sigma}}^{\dagger}
                   c_{2 \sigma}^{\dagger}b_{2 \bar{\sigma}}^{\dagger}
                 + c_{1 \sigma}b_{1 \bar{\sigma}}
                   c_{2 \sigma}b_{2 \bar{\sigma}} 
               \right) ,
\end{align}
where $\bar{\sigma}$ in eq.~(\ref{eq:H_{FB}}) means
$\bar{\uparrow} = \downarrow$ and $\bar{\downarrow} = \uparrow$.
The simple replacement $1 \to 3$ in the expression of $H_{12}$
yields $H_{23}$.
In terms of the vacuum state $|0\rangle$,
the average and second moment of the transmission probability
$\mathcal{T}_{ij}$ are expressed as~\cite{mathur}
\begin{align}
   \langle \mathcal{T}_{ij} \rangle
 & = \langle 0 |c_{i \downarrow}c_{i \uparrow} {\rm e}^{- \tilde{D}_{L} H}
   c_{j \uparrow}^{\dagger}c_{j \downarrow}^{\dagger} | 0 \rangle ,
           \\
       \label{eq:T^2-expression}
   \langle \mathcal{T}_{ij}^{2} \rangle
 & = \langle 0 |c_{i \downarrow}c_{i \uparrow}b_{i \downarrow}b_{i \uparrow}
     {\rm e}^{- \tilde{D}_{L} H}
     b_{j \uparrow}^{\dagger}b_{j \downarrow}^{\dagger}
     c_{j \uparrow}^{\dagger}c_{j \downarrow}^{\dagger} | 0 \rangle
\end{align}
with $\tilde{D}_{L} = D L$.
Here, $c_{i \downarrow}c_{i \uparrow}$, $b_{i \downarrow}b_{i \uparrow}$
and their Hermitian conjugates play a role of the current vertex,
where $\uparrow$ and $\downarrow$ correspond to the retarded and advanced
sectors, respectively.
We see that $H$ is non-Hermitian because
${H_{12}^{\rm FB}}^{\dagger} = - H_{12}^{\rm FB}$ and
${H_{23}^{\rm FB}}^{\dagger} = - H_{23}^{\rm FB}$,
but it has only real eigenvalues.
Equation~(\ref{eq:T^2-expression}) has not been presented
in ref.~\citen{mathur},
but we can easily derive it by extending Mathur's argument.

Our task is now reduced to evaluating the correlation functions
for the non-random interacting fermion-boson system.
Note that $H_{12}^{\rm FB}$ and $H_{23}^{\rm FB}$ annihilate
the two-fermion state
$c_{2 \uparrow}^{\dagger}c_{2 \downarrow}^{\dagger} | 0 \rangle$,
while
$H_{12}^{\rm F}c_{2 \uparrow}^{\dagger}c_{2 \downarrow}^{\dagger} | 0 \rangle
= H_{23}^{\rm F}c_{2 \uparrow}^{\dagger}c_{2 \downarrow}^{\dagger} | 0 \rangle
= (1/2)c_{2 \uparrow}^{\dagger}c_{2 \downarrow}^{\dagger} | 0 \rangle$.
Consequently, the expressions for
$\langle {g'}_{m} \rangle = \langle \mathcal{T}_{22} \rangle$ and
$\langle {g'}_{m}^{2} \rangle = \langle \mathcal{T}_{22}^{2} \rangle$
are simplified to
\begin{align}
      \label{eq:g-expression}
  \langle {g'}_{m} \rangle
  & = \langle 0 | {\rm e}^{- \tilde{D}_{L} H_{m}^{\rm B}}| 0 \rangle ,
          \\
      \label{eq:g^2-expression}
  \langle {g'}_{m}^{2} \rangle
  & = \langle 0 |b_{2 \downarrow}b_{2 \uparrow}
      {\rm e}^{- \tilde{D}_{L} H_{m}^{\rm B}}
      b_{2 \uparrow}^{\dagger}b_{2 \downarrow}^{\dagger}| 0 \rangle
\end{align}
with $H_{0}^{\rm B} \equiv H_{12}^{\rm B} + 1/2$
and $H_{1}^{\rm B} \equiv H_{12}^{\rm B} + H_{23}^{\rm B} + 1$.
Equations~(\ref{eq:g-expression}) and (\ref{eq:g^2-expression}) indicate that
we can evaluate $\langle {g'}_{m} \rangle$ and $\langle {g'}_{m}^{2} \rangle$
in the asymptotic regime of $L \gg D^{-1}$ if the low-energy eigenstates of
the boson Hamiltonian $H_{m}^{\rm B}$ ($m = 1, 2$) are obtained.
That is, we need not consider the fermion degrees of freedom
in the following argument.
This enables us to treat our task in an analytical manner.
It should be emphasized that
this special simplification arises only in the case that
we treat the right-moving conductance $g'$ with $M = 2$ or $3$.
In other words, an analytical treatment seems to be difficult for $M \ge 4$.

To evaluate $\langle {g'}_{0}^{2} \rangle$,
we consider eigenstates $| \Psi \rangle$ of $H_{0}^{\rm B}$.
We need to obtain the eigenstates having an overlap
with $b_{2 \uparrow}^{\dagger}b_{2 \downarrow}^{\dagger}| 0 \rangle$,
and therefore we focus on the Hilbert space spanned by
\begin{align}
  | \psi^{n} \rangle
    = \frac{1}{n!(n+1)!}
      \left(b_{1\uparrow}^{\dagger}b_{1\downarrow}^{\dagger}\right)^{n}
      \left(b_{2\uparrow}^{\dagger}b_{2\downarrow}^{\dagger}\right)^{n+1}
      | 0 \rangle
\end{align}
with $n = 0, 1,2,\dots, \infty$.
Eigenstates of $H_{0}^{\rm B}$ is expressed as
$| \Psi \rangle = \sum_{n=0}^{\infty} c_{n} | \psi^{n} \rangle$.
From the eigenvalue equation $H_{0}^{\rm B}| \Psi \rangle = E | \Psi \rangle$,
we obtain the recurrence relation
\begin{align}
      \label{eq:recurrence1}
  (2n^{2} + 4n + 2 - E) c_{n} - n(n+1) c_{n-1} - (n+1)(n+2) c_{n+1} = 0 .
\end{align}
On the basis of eq.~(\ref{eq:recurrence1}),
we obtain the continuous eigenstates $| k \rangle$ which satisfy
$H_{0}^{\rm B} | k \rangle = \frac{1}{4}(1+k^{2})| k \rangle$ and
$\langle 0 |b_{2 \downarrow}b_{2 \uparrow}| k \rangle = 1$ with $k > 0$.
The orthogonality condition is
\begin{align}
     \label{eq:orthonomalization1}
  \langle k' | k \rangle 
   = \frac{8 \cosh \left(\frac{\pi k}{2}\right)}
          {\pi k \sinh \left(\frac{\pi k}{2}\right)}
     \delta\left( k - k' \right) .
\end{align}
The derivation of the above result is briefly described in Appendix B.
From eq.~(\ref{eq:orthonomalization1}),
we observe that the identity operator $\mathcal{I}$ is expressed as
\begin{align}
     \label{eq:identity1}
  \mathcal{I}
   = \int_{0}^{\infty} {\rm d}k
     \frac{\pi k \sinh \left( \frac{\pi k}{2} \right)}
          {8 \cosh \left( \frac{\pi k}{2} \right)}
     | k \rangle \langle k | .
\end{align}
We now evaluate $\langle {g'}_{0}^{2} \rangle$.
Inserting eq.~(\ref{eq:identity1}) into eq.~(\ref{eq:g^2-expression}),
we obtain
\begin{align}
  \langle {g'}_{0}^{2} \rangle
    = \int_{0}^{\infty} {\rm d}k
      \frac{\pi k \sinh \left( \frac{\pi k}{2} \right)}
           {8 \cosh \left( \frac{\pi k}{2} \right)}
      \left|\langle 0 | b_{2 \downarrow}b_{2 \uparrow}| k \rangle \right|^{2}
      {\rm e}^{- \tilde{D}_{L} \frac{1+k^{2}}{4}} .
\end{align}
The main contribution to the $k$-integration comes from the region of
$k \ll 1$, in which we can approximate as
\begin{align}
 \frac{\pi k \sinh \left( \frac{\pi k}{2} \right)}
           {8 \cosh \left( \frac{\pi k}{2} \right)}
  = \frac{\pi^{2}}{16} k^{2} .
\end{align}
After the $k$-integration, we obtain
\begin{align}
  \langle {g'}_{0}^{2} \rangle
    = \frac{\pi^{\frac{5}{2}}}{64} \frac{1}{\sqrt{\frac{DL}{4}}^{3}}
      {\rm e}^{- \frac{DL}{4}} .
\end{align}
This result is identical to eq.~(\ref{eq:g^2_m=0}) in the case of
$N = 1$ under the replacement $DL \to s$.

To evaluate $\langle {g'}_{1} \rangle$  in the asymptotic regime,
we consider low-lying eigenstates $| \Psi \rangle$ of
$H_{1}^{\rm B}$ having an overlap with $| 0 \rangle$.~\cite{takane5}
Therefore, we focus on the Hilbert space spanned by
\begin{align}
  | \psi^{n,m} \rangle
    = \frac{1}{n!(n+m)!m!}
      \left(b_{1\uparrow}^{\dagger}b_{1\downarrow}^{\dagger}\right)^{n}
      \left(b_{2\uparrow}^{\dagger}b_{2\downarrow}^{\dagger}\right)^{n+m}
      \left(b_{3\uparrow}^{\dagger}b_{3\downarrow}^{\dagger}\right)^{m}
      | 0 \rangle
\end{align}
with $n, m = 0, 1,2,\dots, \infty$.
Eigenstates of $H_{1}^{\rm B}$ is expressed as
$| \Psi \rangle = \sum_{n,m=0}^{\infty} c^{n,m} | \psi^{n,m} \rangle$.
From the eigenvalue equation $H_{1}^{\rm B}| \Psi \rangle = E | \Psi \rangle$,
we obtain the recurrence relation
\begin{align}
      \label{eq:recurrence2'}
  & \left\{ 1 + 2(n+m+1)(n+m+1/2) - E \right\} c^{n,m}
     - n(n+m) c^{n-1,m} - (n+m)m c^{n,m-1}
  \nonumber \\
  & \hspace{20mm}
     - (n+1)(n+m+1) c^{n+1,m} - (n+m+1)(m+1) c^{n,m+1} = 0 .
\end{align}
We restrict our attention to the lowest energy branch
of the excitation spectrum, which determines
the asymptotic behavior of $\langle {g'}_{1} \rangle$.
We assume for this branch that $c^{n,m}$ depends on only $n + m$
(i.e., $c^{n+m,0} = \cdots = c^{n,m} = \dots = c^{0,n+m}$),
which has been confirmed in ref.~\citen{takane5}
by a numerical diagonalization of eq.~(\ref{eq:recurrence2'}).
Under this assumption, we rewrite $c^{n,m}$ as $c^{n,m} \to c_{l}$
with $l \equiv n + m$.
Equation~(\ref{eq:recurrence2'}) is then reduced to
\begin{align}
      \label{eq:recurrence2}
    \left\{ 1 + 2(l+1)(l+1/2) - E \right\} c_{l}
     - l^{2} c_{l-1} - (l+1)(l+2) c_{l+1} = 0 .
\end{align}
On the basis of eq.~(\ref{eq:recurrence2}),
we obtain the continuous eigenstates
$| k \rangle$ which satisfy $H_{1}^{\rm B} | k \rangle = (1+k^{2})| k \rangle$
and $\langle 0 | k \rangle = 1$ with $k > 0$.
The orthogonality condition is
\begin{align}
     \label{eq:orthonomalization2}
  \langle k' | k \rangle 
   = \frac{\sinh^{2}(\pi k)}{2\pi k^{2} \cosh (\pi k)}
     \delta\left( k - k' \right) .
\end{align}
The derivation of the above result is briefly described in Appendix C.
From eq.~(\ref{eq:orthonomalization2}),
we observe that the identity operator $\mathcal{I}$
in the restricted Hilbert space is expressed as
\begin{align}
     \label{eq:identity2}
  \mathcal{I}
   = \int_{0}^{\infty} {\rm d}k
     \frac{2\pi k^{2} \cosh (\pi k)}{\sinh^{2} (\pi k)}
     | k \rangle \langle k | .
\end{align}
Inserting eq.~(\ref{eq:identity2}) into eq.~(\ref{eq:g-expression}),
we obtain
\begin{align}
  \langle {g'}_{1} \rangle
    = \int_{0}^{\infty} {\rm d}k
      \frac{2 \pi k^{2} \cosh (\pi k)}{\sinh^{2} (\pi k)}
      \left|\langle 0 | k \rangle \right|^{2}
      {\rm e}^{- \tilde{D}_{L} (1+k^{2})} .
\end{align}
Carrying out the $k$-integration, we obtain
\begin{align}
  \langle {g'}_{1} \rangle
    = \frac{1}{2\sqrt{\pi}} \frac{1}{\sqrt{\frac{DL}{4}}}
      {\rm e}^{- DL} .
\end{align}
This result is identical to eq.~(\ref{eq:g_m=1}) in the case of
$N = 1$ under the replacement $DL \to s$.

Finally, we evaluate $\langle {g'}_{1}^{2} \rangle$  in the asymptotic regime.
To do so, we consider low-lying eigenstates $| \Phi \rangle$
of $H_{1}^{\rm B}$ having an overlap
with $b_{2 \uparrow}^{\dagger}b_{2 \downarrow}^{\dagger}| 0 \rangle$.
Therefore, we focus on the Hilbert space spanned by
\begin{align}
  | \phi^{n,m} \rangle
    = \frac{1}{n!(n+m+1)!m!}
      \left(b_{1\uparrow}^{\dagger}b_{1\downarrow}^{\dagger}\right)^{n}
      \left(b_{2\uparrow}^{\dagger}b_{2\downarrow}^{\dagger}\right)^{n+m+1}
      \left(b_{3\uparrow}^{\dagger}b_{3\downarrow}^{\dagger}\right)^{m}
      | 0 \rangle
\end{align}
with $n, m = 0, 1,2,\dots, \infty$.
Eigenstates of $H_{1}^{\rm B}$ are expressed as
$| \Phi \rangle = \sum_{n,m=0}^{\infty} d^{n,m} | \phi^{n,m} \rangle$.
From the eigenvalue equation $H_{1}^{\rm B}| \Phi \rangle = E | \Phi \rangle$,
we obtain the recurrence relation
\begin{align}
      \label{eq:recurrence3'}
  & \left\{ 1 + 2(n+m+1)(n+m+3/2) - E \right\} d^{n,m}
     - n(n+m+1) d^{n-1,m} - (n+m+1)m d^{n,m-1}
  \nonumber \\
  & \hspace{20mm}
     - (n+1)(n+m+2) d^{n+1,m} - (n+m+2)(m+1) d^{n,m+1} = 0 .
\end{align}
Again, our attention is restricted to the lowest energy branch
of the excitation spectrum.
We assume for this branch that $d^{n,m}$ depends on only $n + m$
(i.e., $d^{n+m,0} = \cdots = d^{n,m} = \dots = d^{0,n+m}$),
which can be confirmed by a numerical diagonalization of
eq.~(\ref{eq:recurrence3'}).
Under this assumption, we rewrite $d^{n,m}$ as $d^{n,m} \to d_{l}$
with $l \equiv n + m$.
Equation~(\ref{eq:recurrence3'}) is then reduced to
\begin{align}
      \label{eq:recurrence3}
    \left\{ 1 + 2(l+1)(l+3/2) - E \right\} d_{l}
     - l(l+1) d_{l-1} - (l+2)^{2} d_{l+1} = 0 .
\end{align}
On the basis of eq.~(\ref{eq:recurrence3}),
we obtain the continuous eigenstates $| k \rangle$
which satisfy $H_{1}^{\rm B} | k \rangle = (1+k^{2})| k \rangle$ and
$\langle 0 |b_{2 \downarrow}b_{2 \uparrow}| k \rangle = -k^{2}$ with $k > 0$.
The orthogonality condition is
\begin{align}
      \label{eq:orthonomalization3}
  \langle k' | k \rangle 
   = \frac{\sinh^{2}(\pi k)}{2\pi \cosh (\pi k)}
     \delta\left( k - k' \right) .
\end{align}
The derivation of the above result is briefly described in Appendix C.
From eq.~(\ref{eq:orthonomalization3}),
we observe that the identity operator $\mathcal{I}$
in the restricted Hilbert space is expressed as
\begin{align}
     \label{eq:identity3}
  \mathcal{I}
   = \int_{0}^{\infty} {\rm d}k
     \frac{2 \pi \cosh(\pi k)}{\sinh^{2}(\pi k)} | k \rangle \langle k | .
\end{align}
Inserting eq.~(\ref{eq:identity3}) into eq.~(\ref{eq:g^2-expression}),
we obtain
\begin{align}
  \langle {g'}_{1}^{2} \rangle
    = \int_{0}^{\infty} {\rm d}k
      \frac{2 \pi \cosh(\pi k)}{\sinh^{2}(\pi k)}
      \left|\langle 0 |b_{2 \downarrow}b_{2 \uparrow}
                  | k \rangle \right|^{2}
      {\rm e}^{- \tilde{D}_{L} (1+k^{2})} .
\end{align}
Carrying out the $k$-integration, we obtain
\begin{align}
  \langle {g'}_{1}^{2} \rangle
    = \frac{1}{16\sqrt{\pi}} \frac{1}{\sqrt{\frac{DL}{4}}^{3}}
      {\rm e}^{- DL} .
\end{align}
This result is identical to eq.~(\ref{eq:g^2_m=1}) in the case of
$N = 1$ under the replacement $DL \to s$.

\section{Summary}

We have studied electron transport properties in disordered unitary wires
of length $L$ in the presence of the channel-number imbalance
between two propagating directions.
Our attention is focused on the case in which the number of
left-moving channels is by $m$ greater than that of the right-moving ones.
In this case, $m$ left-moving channels become perfectly conducting and
the dimensionless conductances $g$ and $g'$ for the left-moving
and right-moving channels, respectively, satisfy $g = g' + m$.
First, we have obtained the average $\langle g' \rangle$ and second moment
$\langle {g'}^{2} \rangle$ of $g' = g - m$ in the long-$L$ regime
by using the exact solution of the DMPK equation.
Both $\langle g' \rangle$ and $\langle {g'}^{2} \rangle$
decay exponentially as a function of $L$.
It is shown that their exponential decay becomes faster with increasing $m$.
This behavior can be understood from the fact that
the eigenvalue repulsion arising from the perfectly conducting eigenvalue
is enhanced with increasing $m$.
Second, we have employed the $M$-edge Chalker-Coddington model for the cases of
$M =2$ and $M = 3$, and obtained $\langle g' \rangle$ and
$\langle {g'}^{2} \rangle$ in the long-$L$ regime
by using the supersymmetry approach.
The case  of $M = 3$ corresponds to the simplest nontrivial example of
the channel-number-imbalanced unitary class.
We have shown that the resulting asymptotic forms of $\langle g' \rangle$ and
$\langle {g'}^{2} \rangle$ are identical to
those obtained from the DMPK equation including the pre-exponential factor.

\appendix
\section{Derivation of the Scaling Relation}

The average of a function $F(\{\lambda_{a}\})$ is defined as
\begin{align}
  \langle F \rangle
 = \int_{0}^{\infty} {\rm d}\lambda_{1} \cdots \lambda_{N}
   F(\{\lambda_{a}\})P(\{\lambda_{a}\},s) .
\end{align}
Using the DMPK equation given in eq.~(\ref{eq:dmpk-lambda}),
we find that~\cite{mello2}
\begin{align}
 N \frac{\partial \langle F \rangle}{\partial s}
 & = \left\langle
          \sum_{a=1}^{N}\frac{1}{J}\frac{\partial}{\partial \lambda_{a}}
          \left\{ \lambda_{a}(1+\lambda_{a})J
                  \frac{\partial F}{\partial \lambda_{a}}
          \right\}
     \right\rangle
              \nonumber \\
 & = \Biggl\langle
        \sum_{a=1}^{N}
        \biggl\{ \lambda_{a}(1+\lambda_{a})
                 \frac{\partial^{2}F}{\partial \lambda_{a}^{2}}
               + (1+2\lambda_{a})\frac{\partial F}{\partial \lambda_{a}}
             \nonumber \\
 & \hspace{25mm}
               + \lambda_{a}(1+\lambda_{a})
                 \Biggl( \sum_{\scriptstyle b=1 \atop \scriptstyle (b \neq a)}
                                              ^{N}
                         \frac{2}{\lambda_{a}-\lambda_{b}}
                       + \frac{m}{\lambda_{a}}
                 \Biggr)\frac{\partial F}{\partial \lambda_{a}}
        \biggr\}
     \Biggr\rangle .
\end{align}
Replacing $F$ by $g' = \sum_{a=1}^{N}(1+\lambda_{a})^{-1}$
in the above equation,
we obtain eq.~(\ref{eq:scaling}) after straightforward calculations.

\appendix
\section{Energy Spectrum of $H_{0}^{\rm B}$}

As shown in the text, eigenstates of $H_{0}^{\rm B}$ satisfying
$\langle 0 |b_{2 \downarrow}b_{2 \uparrow}| \Psi \rangle \neq 0$
are expressed as
\begin{align}
  | \Psi (E) \rangle = \sum_{n=0}^{\infty} c_{n}(E) | \psi^{n} \rangle ,
\end{align}
where $c_{n}$ satisfies eq.~(\ref{eq:recurrence1}).
We obtain the asymptotic form of $c_{n}$
in the large-$n$ limit adapting the manipulation given by Mathur.~\cite{mathur}
We introduce the generating function defined by
$f(x) = \sum_{n = 0}^{\infty} c_{n}x^{n}$.
Using eq.~(\ref{eq:recurrence1}), we can show that $f(x)$ obeys
\begin{align}
 x(x-1)^{2}\frac{d^{2}f}{dx^{2}} + 2(2x-1)(x-1)\frac{df}{dx}
    + (2x - 2 + E)f = 0 .
\end{align}
It is convenient to rewrite $f(x)$ as $f(x) = (1-x)^{\mu}g(x)$.
If we set
\begin{align}
   \mu = - \frac{1}{2} + \kappa
\end{align}
with $\kappa = {\rm i} \sqrt{4E-1}/2$ for $E > 1/4$ and
$\kappa = \sqrt{1-4E}/2$ for $E < 1/4$,
then $g(x)$ is expressed in terms of the hypergeometric function as
$g(x) = F(2+\mu,1+\mu,2;x)$.
The coefficient $c_{n}$ is expressed as
\begin{align}
     \label{eq:extraction}
  c_{n} = \frac{1}{2\pi {\rm i}} \int_{C} {\rm d}x
          \frac{1}{x^{n+1}}(1-x)^{\mu}g(x) ,
\end{align}
where $C$ denotes a small contour encircling the origin
in the anticlockwise direction and $c_{0} = 1$ has been assumed.
We employ the integral representation of the hypergeometric function
\begin{align}
     \label{eq:integral-rep}
  F(a,b,c;x)
  = \frac{\Gamma(c)}{\Gamma(b)\Gamma(c-b)}
    \int_{1}^{\infty} {\rm d}t (t-x)^{-a}t^{a-c}(t-1)^{c-b-1} ,
\end{align}
which is justified when ${\rm Re} [c] > {\rm Re} [b] > 0$.
Substituting this into eq.~(\ref{eq:extraction}) and exchanging
the order of the integrations over $x$ and $t$, we obtain
\begin{align}
  c_{n} = \frac{1}{\Gamma(1+\mu)\Gamma(1-\mu)}
          \int_{1}^{\infty} {\rm d}t \, t^{\mu}(t-1)^{-\mu}
          \int_{C} \frac{{\rm d}x}{2\pi {\rm i}}
          \frac{1}{x^{n+1}} (t-x)^{-2-\mu} (1-x)^{\mu} .
\end{align}
We here draw the branch cut between $x = 1$ and $x = t$,
and take the phase of $(t-x)$ and $(1-x)$ being zero
when $x$ lies on the real axis to the left of $1$.
Deforming the contour $C$ as in ref.~\citen{mathur},
we can show that $C$ is replaced by the contour starting from $x = 1$
to $x = t$ above the branch cut and after encircling the point $x = t$
in the clockwise direction, coming back to $x = 1$ below the branch cut.
Exchanging the order of the integrations, we obtain
\begin{align}
  c_{n} = - \frac{1}{\Gamma(1+\mu)\Gamma(1-\mu)}
          \int_{1}^{\infty} {\rm d}x \frac{(x-1)^{\mu}}{x^{n+1}}
          \int_{C_{x}} \frac{{\rm d}t}{2\pi {\rm i}}
          \, t^{\mu}(t-1)^{-\mu}(x-t)^{-2-\mu} ,
\end{align}
where we have drawn the branch cut between $t = x$ and $t = \infty$
on the real axis and $C_{x}$ denotes the contour coming from $t = \infty$ to
$t = x$ below the branch cut and after encircling the point $t = x$
in the clockwise direction, going back to $t = \infty$ above the branch cut.
Rescaling $t$ as $u \equiv t/x$, we obtain
\begin{align}
  c_{n} = - \frac{1}{\Gamma(1+\mu)\Gamma(1-\mu)}
          \int_{1}^{\infty} {\rm d}x \frac{(x-1)^{\mu}}{x^{n+2+\mu}}
          \sigma(x)
\end{align}
with
\begin{align}
  \sigma(x) = \int_{C_{1}} \frac{{\rm d}u}{2\pi {\rm i}}
              \, u^{\mu}(u-\frac{1}{x})^{-\mu}(1-u)^{-2-\mu} ,
\end{align}
where $C_{1}$ is identical to $C_{x}$ with $x \to 1$.
Using an analytic continuation, we can show that
\begin{align}
  \sigma(x) = \frac{\mu}{x}F\left(1+\mu,2+\mu,2;\frac{1}{x}\right) .
\end{align}
Then, we obtain
\begin{align}
  c_{n} = \frac{1}{\Gamma(\frac{1}{2}+\kappa)\Gamma(\frac{1}{2}-\kappa)}
          \int_{1}^{\infty} {\rm d}x \frac{(x-1)^{-\frac{1}{2}+\kappa}}
                                          {x^{n+\frac{5}{2}+\kappa}}
          F\left(\frac{1}{2}+\kappa,\frac{3}{2}+\kappa,2;\frac{1}{x}\right) ,
\end{align}
where $\mu = -1/2 + \kappa$ has been used.
The change of the variable from $x$ to $s \equiv \ln x$ results in
\begin{align}
  c_{n} = \frac{1}{\Gamma(\frac{1}{2}+\kappa)\Gamma(\frac{1}{2}-\kappa)}
          \int_{0}^{\infty} {\rm d}s
          ({\rm e}^{s}-1)^{-\frac{1}{2}+\kappa}
          {\rm e}^{-\left( n+\frac{3}{2}+\kappa \right)s}
          F\left(\frac{1}{2}+\kappa,\frac{3}{2}+\kappa,2;{\rm e}^{-s}\right) .
\end{align}
The above equation indicates that the behavior of the integrand for
$s \approx 0$ is important in considering the large-$n$ limit.
Therefore, we employ the approximations
$({\rm e}^{s}-1)^{-\frac{1}{2}+\kappa} \approx s^{-\frac{1}{2}+\kappa}$ and
\begin{align}
      \label{eq:approx-F}
  F(a,b,c;{\rm e}^{-s}) \approx  \frac{\Gamma(c)\Gamma(c-a-b)}
                           {\Gamma(c-a)\Gamma(c-b)}
                    + \frac{\Gamma(c)\Gamma(a+b-c)}
                           {\Gamma(a)\Gamma(b)} (1-{\rm e}^{-s})^{c-a-b}
\end{align}
which are applicable when ${\rm e}^{-s} \approx 1$.
Carrying out the $s$-integration, we obtain
\begin{align}
  c_{n} = \frac{1}{\Gamma(\frac{1}{2}+\kappa)\Gamma(\frac{1}{2}-\kappa)}
          \left( \frac{\Gamma(2\kappa)\Gamma(\frac{1}{2}-\kappa)}
                     {\Gamma(\frac{1}{2}+\kappa)\Gamma(\frac{3}{2}+\kappa)}
                 \frac{{\rm e}^{\kappa \ln n}}{\sqrt{n}}
               + \frac{\Gamma(-2\kappa)\Gamma(\frac{1}{2}+\kappa)}
                     {\Gamma(\frac{1}{2}-\kappa)\Gamma(\frac{3}{2}-\kappa)}
                 \frac{{\rm e}^{-\kappa \ln n}}{\sqrt{n}}
          \right) .
\end{align}

If $E > 1/4$, we can set $\kappa = {\rm i}k/2$ with $k = \sqrt{4E-1}$.
In this case, $c_{n}$ in the large-$n$ limit is expressed as
\begin{align}
      \label{eq:c_n-H0}
  c_{n}(E) = \frac{\alpha_{k}}{\sqrt{n}}
             \cos \left(\frac{k}{2}\ln n + \zeta_{k} \right)
\end{align}
with
\begin{align}
  \alpha_{k}
  & = \frac{4\cosh\left(\frac{\pi k}{2}\right)}{\pi \sqrt{k\sinh(\pi k)}} ,
               \\
  \zeta_{k}
  & = {\rm arg}
      \left( \frac{\Gamma({\rm i}k)\Gamma(\frac{1}{2}-{\rm i}\frac{k}{2})}
                  {\Gamma(\frac{1}{2}+{\rm i}\frac{k}{2})
                  \Gamma(\frac{3}{2}+{\rm i}\frac{k}{2})}
      \right) .
\end{align}
We show that the orthonormalization of $| \Psi (E) \rangle$
is possible when $E > 1/4$.
Note that $\langle \Psi (E') | \Psi (E) \rangle
= \sum_{n=0}^{\infty}c_{n}(E')c_{n}(E)$.
Using eq.~(\ref{eq:recurrence1}), we can express the partial sum
$\sum_{n=0}^{M}c_{n}(E')c_{n}(E)$ as
\begin{align}
    \label{eq:sum-to-M}
  \sum_{n=0}^{M}c_{n}(E')c_{n}(E)
    = \frac{(M+1)(M+2)}{E-E'}
      \left( c_{M+1}(E')c_{M}(E) - c_{M}(E')c_{M+1}(E) \right) .
\end{align}
Substituting eq.~(\ref{eq:c_n-H0}) into eq.~(\ref{eq:sum-to-M})
and then taking the limit of $M \to \infty$, we obtain
\begin{align}
    \label{eq:sum-to-infty}
  \sum_{n=0}^{\infty}c_{n}(E')c_{n}(E)
   = \pi \alpha_{k}^{2} \delta\left( k - k' \right) .
\end{align}
This indicates that $| \Psi (E) \rangle$ can be orthonormalized when $E > 1/4$.
In contrast, when $E < 1/4$, the partial sum $\sum_{n=0}^{M}c_{n}(E')c_{n}(E)$
does not converges in the large-$M$ limit even if $E \neq E'$.
Hence, the orthonormalization is impossible.
We conclude that $| \Psi (E) \rangle$ exits only when $E > 1/4$.
It is convenient to use $k$ instead of $E$.
We rewrite the eigenstate as $| \Psi (E) \rangle \to | k \rangle$
with $E = \frac{1}{4}(1 + k^{2})$.
Equation~(\ref{eq:sum-to-infty}) is then rewritten as
\begin{align}
  \langle k' | k \rangle 
   = \frac{8 \cosh \left(\frac{\pi k}{2}\right)}
          {\pi k \sinh \left(\frac{\pi k}{2}\right)}
     \delta\left( k - k' \right) .
\end{align}
Finally, we note that $c_{0} = 1$ means
$\langle 0 |b_{2 \downarrow}b_{2 \uparrow}| k \rangle = 1$.

\appendix
\section{Low Energy Spectrum of $H_{1}^{\rm B}$}

We first consider the low-lying eigenstates of $H_{1}^{\rm B}$ satisfying
$\langle 0 | \Psi \rangle \neq 0$.
Our attention is restricted to the lowest energy branch in which
the eigenstates are expressed as
\begin{align}
  | \Psi (E) \rangle = \sum_{n,m=0}^{\infty} c_{n+m}(E) | \psi^{n,m} \rangle ,
\end{align}
where $c_{l}$ with $l = n + m$ satisfies eq.~(\ref{eq:recurrence2}).
We obtain the asymptotic form of $c_{l}$ in the large-$l$ limit.
We introduce the generating function defined by
$f(x) = \sum_{l = 0}^{\infty} c_{l}x^{l}$.
Using eq.~(\ref{eq:recurrence2}), we can show that $f(x)$ obeys
\begin{align}
 x(x-1)^{2}\frac{d^{2}f}{dx^{2}} + (3x-2)(x-1)\frac{df}{dx}
    + (x - 2 + E)f = 0 .
\end{align}
It is convenient to rewrite $f(x)$ as $f(x) = (1-x)^{\mu}g(x)$.
If we set $\mu = {\rm i} \sqrt{E-1}$ for $E > 1$ and
$\mu = \sqrt{1-E}$ for $E < 1$,
then $g(x)$ is expressed in terms of the hypergeometric function as
$g(x) = F(1+\mu,1+\mu,2;x)$.
The coefficient $c_{l}$ is expressed as
\begin{align}
     \label{eq:extraction'}
  c_{l} = \frac{1}{2\pi {\rm i}} \int_{C} {\rm d}x
          \frac{1}{x^{l+1}}(1-x)^{\mu}g(x) ,
\end{align}
where $C$ denotes a small contour encircling the origin
and $c_{0} = 1$ has been assumed.
Using eq.~(\ref{eq:integral-rep}) and
adapting the procedure described in Appendix B, we obtain
\begin{align}
  c_{l} = \frac{\sin(\pi \mu)}{\pi \mu}
          \int_{1}^{\infty} {\rm d}x \frac{(x-1)^{\mu}}{x^{l+\mu+2}}
          F\left(\mu,1+\mu,1;\frac{1}{x}\right) .
\end{align}
After changing the variable from $x$ to $s = \ln x$,
we replace the hypergeometric function by the approximate expression
given in eq.~(\ref{eq:approx-F}).
Carrying out the $s$-integration, we obtain
\begin{align}
      \label{eq:asymptotic}
 c_{l} =
   \frac{\sin(\pi \mu)}{\pi \mu}
   \left( \frac{\Gamma(1-\mu)\Gamma(2\mu)}{\Gamma(1+\mu)\Gamma(\mu)}
          \frac{{\rm e}^{\mu \ln (l+1)}}{l+1}
        + \frac{\Gamma(1+\mu)\Gamma(-2\mu)}{\Gamma(1-\mu)\Gamma(-\mu)}
          \frac{{\rm e}^{-\mu \ln (l+1)}}{l+1}
   \right) .
\end{align}

When $E > 1$, we write $\mu = {\rm i} k$ with $k \equiv \sqrt{E - 1}$.
In this case, $c_{l}$ in the large-$l$ limit is expressed as
\begin{align}
      \label{eq:c_l-H1}
 c_{l}(E) = \frac{\beta_{k}}{l+1} \cos(k\ln (l+1) + \eta_{k})
\end{align}
with
\begin{align}
     \label{eq:beta}
  \beta_{k}
  & = \frac{\sinh (\pi k)}{\pi k \sqrt{\cosh(\pi k)}} ,
               \\
  \eta_{k}
  & = {\rm arg}
      \left( \frac{\Gamma(1-{\rm i}k)\Gamma(2{\rm i}k)}
                  {\Gamma(1+{\rm i}k)\Gamma({\rm i}k)}
      \right) .
\end{align}
We show that the orthonormalization of $| \Psi (E) \rangle$
is possible when $E > 1$.
Note that
\begin{align}
  \langle \Psi (E') | \Psi (E) \rangle
 & = \sum_{n,m=0}^{\infty}c^{n,m}(E')c^{n,m}(E)
              \nonumber \\
 & = \sum_{l=0}^{\infty}(l+1)c_{l}(E')c_{l}(E) .
\end{align}
Using eq.~(\ref{eq:recurrence2}), we can express the partial sum
$\sum_{l=0}^{M}(l+1)c_{l}(E')c_{l}(E)$ as
\begin{align}
    \label{eq:sum-to-M'}
  \sum_{l=0}^{M}(l+1)c_{l}(E')c_{l}(E)
    = \frac{(M+1)^{2}(M+2)}{E-E'}
      \left( c_{M+1}(E')c_{M}(E) - c_{M}(E')c_{M+1}(E) \right) .
\end{align}
Substituting eq.~(\ref{eq:c_l-H1}) into eq.~(\ref{eq:sum-to-M'})
and then taking the limit of $M \to \infty$, we obtain
\begin{align}
    \label{eq:sum-to-infty'}
  \sum_{l=0}^{\infty}(l+1)c_{l}(E')c_{l}(E)
   = \frac{\pi \beta_{k}^{2}}{2} \delta\left( k - k' \right) .
\end{align}
This indicates that $| \Psi (E) \rangle$ can be orthonormalized when $E > 1$.
In contrast, when $E < 1$, we can show that the partial sum diverges
in the limit of $M \to \infty$ even if $E \neq E'$,
and thereby the orthonormalization is impossible.
We conclude that $| \Psi (E) \rangle$ exists only when $E > 1$.
It is convenient to rewrite the eigenstate as
$| \Psi (E) \rangle \to | k \rangle$ with $E = 1 + k^{2}$.
Equation~(\ref{eq:sum-to-infty'}) is rewritten as
\begin{align}
  \langle k' | k \rangle 
   = \frac{\sinh^{2}(\pi k)}{2\pi k^{2} \cosh (\pi k)}
     \delta\left( k - k' \right) .
\end{align}
We obtain $\langle 0 | k \rangle = 1$ from $c_{0} = 1$.

We next consider the low-lying eigenstates of $H_{1}^{\rm B}$ satisfying
$\langle 0 |b_{2 \downarrow}b_{2 \uparrow}| \Phi \rangle \neq 0$.
We restrict our attention to the lowest energy branch
in which the eigenstates are expressed as
\begin{align}
  | \Phi (E) \rangle = \sum_{n,m=0}^{\infty} d_{n+m}(E) | \phi^{n,m} \rangle ,
\end{align}
where $d_{l}$ with $l = n + m$ satisfies eq.~(\ref{eq:recurrence3}).
We introduce the generating function defined by
$f(x) = \sum_{l = 0}^{\infty} d_{l}x^{l}$.
Using eq.~(\ref{eq:recurrence3}), we can show that $f(x)$ obeys
\begin{align}
 x^{2}(x-1)^{2}\frac{d^{2}f}{dx^{2}} + x(x-1)(4x-3)\frac{df}{dx}
    + \{2x^{2} + (E - 4)x + 1\}f = 0 .
\end{align}
It is convenient to rewrite $f(x)$ as $f(x) = (1-x)^{\mu}x^{-1}g(x)$.
If we set $\mu = {\rm i} \sqrt{E-1}$ for $E > 1$ and
$\mu = \sqrt{1-E}$ for $E < 1$,
then $g(x)$ is expressed in terms of the hypergeometric function as
$g(x) = F(1+\mu,\mu,1;x)$.
The coefficient $c_{l}$ is expressed as
\begin{align}
     \label{eq:extraction''}
  d_{l} = \frac{1}{2\pi {\rm i}} \int_{C} {\rm d}x
            \frac{1}{x^{l+1}}(1-x)^{\mu}\frac{g(x)}{x} ,
\end{align}
where $C$ denotes a small contour encircling the origin.
It should be noted that
eq.~(\ref{eq:extraction''}) indicates $d_{0} = \mu^{2}$.
For $g(x)$, we cannot directly employ the integral representation
of the hypergeometric function given in eq.~(\ref{eq:integral-rep})
when $\mu$ is pure imaginary, so we modify $g(x)$ as
\begin{align}
     g(x) =   F(1+\mu,1+\mu,2;x)
            - \frac{1-\mu^{2}}{2}x F(2+\mu,1+\mu,3;x) .
\end{align}
Now, we can apply eq.~(\ref{eq:integral-rep}) to each term
in the right-hand side of the above equation.
Adapting the procedure described in Appendix B, we obtain
\begin{align}
  d_{l}
   & = \frac{\sin(\pi \mu)}{\pi \mu}
       \int_{1}^{\infty} {\rm d}x
       \Biggl(  \frac{(x-1)^{\mu}}{x^{l+3+\mu}}
                F\left(\mu,1+\mu,1;\frac{1}{x}\right)
                   \nonumber \\
   & \hspace{10mm}
              + (1+\mu)\frac{(x-1)^{\mu}}{x^{l+2+\mu}}
                \left\{  F\left(-1+\mu,2+\mu,1;\frac{1}{x}\right)
                      - F\left(-1+\mu,1+\mu,1;\frac{1}{x}\right)
                \right\}
          \Biggr) .
\end{align}
After changing the variable from $x$ to $s = \ln x$,
we replace the hypergeometric functions by the approximate expression
given in eq.~(\ref{eq:approx-F}).
Carrying out the $s$-integration, we obtain
\begin{align}
      \label{eq:asymptotic'}
 d_{l} =
   \frac{\sin(\pi \mu)}{2\pi}
   \left( \frac{\Gamma(1+2\mu)\Gamma(1-\mu)}{\Gamma(1+\mu)^{2}}
          \frac{{\rm e}^{\mu \ln (l+1)}}{l+1}
        - \frac{\Gamma(1-2\mu)\Gamma(1+\mu)}{\Gamma(1-\mu)^{2}}
          \frac{{\rm e}^{-\mu \ln (l+1)}}{l+1}
   \right) ,
\end{align}
where several higher order terms with respect to $\mu$ are neglected.

When $E > 1$, we write $\mu = {\rm i} k$ with $k \equiv \sqrt{E - 1}$.
In this case, $d_{l}$ in the large-$l$ limit is expressed as
\begin{align}
      \label{eq:d_l-H1}
 d_{l}(E) = \frac{\gamma_{k}}{l+1} \sin(k\ln (l+1) + \theta_{k})
\end{align}
with
\begin{align}
     \label{eq:beta}
  \gamma_{k}
  & = - \frac{\sinh (\pi k)}{\pi \sqrt{\cosh(\pi k)}} ,
               \\
  \theta_{k}
  & = {\rm arg}
      \left( \frac{\Gamma(1+2{\rm i}k)\Gamma(1-{\rm i}k)}
                  {\Gamma(1+{\rm i}k)^{2}}
      \right) .
\end{align}
We show that the orthonormalization of $| \Phi (E) \rangle$
is possible when $E > 1$.
Note that $\langle \Phi (E') | \Phi (E) \rangle
= \sum_{l=0}^{\infty}(l+1)d_{l}(E')d_{l}(E)$.
Using eq.~(\ref{eq:recurrence3}), we can show
\begin{align}
    \label{eq:sum-to-M''}
  \sum_{l=0}^{M}(l+1)d_{l}(E')d_{l}(E)
    = \frac{(M+1)(M+2)^{2}}{E-E'}
      \left( d_{M+1}(E')d_{M}(E) - d_{M}(E')d_{M+1}(E) \right) .
\end{align}
Substituting eq.~(\ref{eq:d_l-H1}) into eq.~(\ref{eq:sum-to-M''})
and then taking the limit of $M \to \infty$, we obtain
\begin{align}
    \label{eq:sum-to-infty''}
  \sum_{l=0}^{\infty}(l+1)d_{l}(E')d_{l}(E)
   = \frac{\pi \gamma_{k}^{2}}{2} \delta\left( k - k' \right) .
\end{align}
This indicates that $| \Phi (E) \rangle$ can be orthonormalized when $E > 1$.
In contrast, when $E < 1$, we can show that the partial sum diverges
in the limit of $M \to \infty$ even if $E \neq E'$,
and thereby the orthonormalization is impossible.
We conclude that $| \Phi (E) \rangle$ exits only when $E > 1$.
It is convenient to rewrite the eigenstate as
$| \Phi (E) \rangle \to | k \rangle$ with $E = 1 + k^{2}$.
Equation~(\ref{eq:sum-to-infty''}) is rewritten as
\begin{align}
  \langle k' | k \rangle 
   = \frac{\sinh^{2}(\pi k)}{2\pi \cosh (\pi k)}
     \delta\left( k - k' \right) .
\end{align}
We obtain $\langle 0 |b_{2 \downarrow}b_{2 \uparrow}| k \rangle = -k^{2}$
from $d_{0} = \mu^{2}$.


\begin{thebibliography}{99}
\bibitem{note} The factor $\pi/2$ is missing in eq.~(57)
of ref.~\citen{mathur}.
\end{thebibliography}
\end{document}